\documentclass[letterpaper,twocolumn,prl,aps,superscriptaddress,amsmath,amssymb,floatfix]{revtex4-2}
\usepackage{mathptmx}
\DeclareMathAlphabet{\mathcal}{OMS}{cmsy}{m}{n}

\usepackage[latin9]{inputenc}
\usepackage{color}
\usepackage{amsmath}
\usepackage{amssymb}
\usepackage{graphicx}
\usepackage{esint}
\usepackage[unicode=true,
 bookmarks=true,bookmarksnumbered=false,bookmarksopen=false,
 breaklinks=false,pdfborder={0 0 1},backref=false,colorlinks=true]
 {hyperref}
\hypersetup{
 linkcolor=magenta,urlcolor=blue,citecolor=blue,pdfstartview={FitH},hyperfootnotes=false}

\makeatletter

\providecommand{\tabularnewline}{\\}

\usepackage{textcomp}
\usepackage{epstopdf}

\pdfpageheight\paperheight
\pdfpagewidth\paperwidth

\providecommand{\tabularnewline}{\\}

\@ifundefined{textcolor}{}{%
 \definecolor{BLACK}{gray}{0}
 \definecolor{WHITE}{gray}{1}
 \definecolor{RED}{rgb}{1,0,0}
 \definecolor{GREEN}{rgb}{0,1,0}
 \definecolor{BLUE}{rgb}{0,0,1}
 \definecolor{CYAN}{cmyk}{1,0,0,0}
 \definecolor{MAGENTA}{cmyk}{0,1,0,0}
 \definecolor{YELLOW}{cmyk}{0,0,1,0}
}

\usepackage{xcolor}\usepackage{soul}
\newcommand{\ket}[1]{\ensuremath{\left|#1\right\rangle}}

\definecolor{blue}{rgb}{0,0,1}
\definecolor{red}{rgb}{1,0,0}
\definecolor{green}{rgb}{0,1,0}

\usepackage{soul}

\makeatother

\begin{document}

\title{Proposal of quantum repeater architecture based on Rydberg atom quantum processors}

\author{Yan-Lei Zhang}
\thanks{These authors contributed equally.}
\affiliation{CAS Key Laboratory of Quantum Information, University of Science
and Technology of China, Hefei, Anhui 230026, China}
\affiliation{CAS Center For Excellence in Quantum Information and Quantum Physics,
University of Science and Technology of China, Hefei, Anhui 230026, China}

\author{Qing-Xuan Jie}
\thanks{These authors contributed equally.}
\affiliation{CAS Key Laboratory of Quantum Information, University of Science
and Technology of China, Hefei, Anhui 230026, China}
\affiliation{CAS Center For Excellence in Quantum Information and Quantum Physics,
University of Science and Technology of China, Hefei, Anhui 230026, China}

\author{Ming Li}
\affiliation{CAS Key Laboratory of Quantum Information, University of Science
and Technology of China, Hefei, Anhui 230026, China}
\affiliation{CAS Center For Excellence in Quantum Information and Quantum Physics,
University of Science and Technology of China, Hefei, Anhui 230026, China}
\affiliation{Hefei National Laboratory, University of Science and Technology of China, Hefei 230088, China}

\author{Shu-Hao Wu}
\affiliation{CAS Key Laboratory of Quantum Information, University of Science
and Technology of China, Hefei, Anhui 230026, China}
\affiliation{CAS Center For Excellence in Quantum Information and Quantum Physics,
University of Science and Technology of China, Hefei, Anhui 230026, China}

\author{Zhu-Bo Wang}
\affiliation{CAS Key Laboratory of Quantum Information, University of Science
and Technology of China, Hefei, Anhui 230026, China}
\affiliation{CAS Center For Excellence in Quantum Information and Quantum Physics,
University of Science and Technology of China, Hefei, Anhui 230026, China}

\author{Xu-Bo Zou}
\affiliation{CAS Key Laboratory of Quantum Information, University of Science
and Technology of China, Hefei, Anhui 230026, China}
\affiliation{CAS Center For Excellence in Quantum Information and Quantum Physics,
University of Science and Technology of China, Hefei, Anhui 230026, China}
\affiliation{Hefei National Laboratory, University of Science and Technology of China, Hefei 230088, China}

\author{Peng-Fei~Zhang}
\affiliation{State Key Laboratory of Quantum Optics and Quantum Optics Devices, and Institute of Opto-Electronics, Shanxi University, Taiyuan 030006, China}
\affiliation{Collaborative Innovation Center of Extreme Optics, Shanxi University, Taiyuan 030006, China.}

\author{Gang Li}
\email{gangli@sxu.edu.cn}
\affiliation{State Key Laboratory of Quantum Optics and Quantum Optics Devices, and Institute of Opto-Electronics, Shanxi University, Taiyuan 030006, China}
\affiliation{Collaborative Innovation Center of Extreme Optics, Shanxi University, Taiyuan 030006, China.}

\author{Tiancai~Zhang}
\affiliation{State Key Laboratory of Quantum Optics and Quantum Optics Devices, and Institute of Opto-Electronics, Shanxi University, Taiyuan 030006, China}
\affiliation{Collaborative Innovation Center of Extreme Optics, Shanxi University, Taiyuan 030006, China.}

\author{Guang-Can Guo}
\affiliation{CAS Key Laboratory of Quantum Information, University of Science
and Technology of China, Hefei, Anhui 230026, China}
\affiliation{CAS Center For Excellence in Quantum Information and Quantum Physics,
University of Science and Technology of China, Hefei, Anhui 230026, China}
\affiliation{Hefei National Laboratory, University of Science and Technology of China, Hefei 230088, China}

\author{Chang-Ling Zou}
\email{clzou321@ustc.edu.cn}
\affiliation{CAS Key Laboratory of Quantum Information, University of Science
and Technology of China, Hefei, Anhui 230026, China}
\affiliation{CAS Center For Excellence in Quantum Information and Quantum Physics,
University of Science and Technology of China, Hefei, Anhui 230026, China}
\affiliation{Hefei National Laboratory, University of Science and Technology of China, Hefei 230088, China}

\date{\today}
\begin{abstract}
Realizing large-scale quantum networks requires the generation of high-fidelity quantum entanglement states between remote quantum nodes, a key resource for quantum communication, distributed computation and sensing applications. However, entanglement distribution between quantum network nodes is hindered by optical transmission loss and local operation errors. Here, we propose a novel quantum repeater architecture that synergistically integrates Rydberg atom quantum processors with optical cavities to overcome these challenges. Our scheme leverages cavity-mediated interactions for efficient remote entanglement generation, followed by Rydberg interaction-based entanglement purification and swapping. Numerical simulations, incorporating realistic experimental parameters, demonstrate the generation of Bell states with 99\% fidelity at rates of 1.1\,kHz between two nodes in local-area network (distance $0.1\,\mathrm{km}$), and can be extend to metropolitan-area ($25\,\mathrm{km}$) or intercity ($\mathrm{250\,\mathrm{km}}$, with the assitance of frequency converters) network with a rate of 0.1\,kHz. This scalable approach opens up near-term opportunities for exploring quantum network applications and investigating the advantages of distributed quantum information processing.
\end{abstract}
\maketitle

\textit{Introduction.-} Recently, great progress have been achieved in the realization of quantum processing units (QPUs) across various platforms, including superconducting circuits~\cite{Arute2019,Wu2021,Xu2023}, trapped ions~\cite{Zhang2017,DeCross2024,Guo2024}, and Rydberg atoms~\cite{Scholl2021,Evered2023,Bluvstein2024}. Among these platforms, the quantum advantages of QPUs have been demonstrated~~\cite{Arute2019,Wu2021,DeCross2024}. To further extend the capability of quantum technologies, a promising direction is the development of quantum networks, where multiple QPUs are interconnected~\cite{Kimble2008,Duan2010,Reiserer2015,Wehner2018,Covey2023}. Quantum networks offer the potential to scale up quantum systems and enabling distributed quantum computing beyond individual QPUs~\cite{Jiang2007,Caleffi2024,Barral2024}. Moreover, they also promises secure communication~\cite{Xu2020,Wei2022}, enhanced sensing~\cite{Gottesman2012,Zhuang2020}, blind quantum computing~\cite{Barz2012}, fundamental studies of quantum entanglement and testing the foundations of quantum mechanics~\cite{Horodecki2009,Hensen2015}. Lies at the heart of quantum networks is the quantum entanglement~\cite{Sangouard2011,Azuma2023}. However, realizing high-fidelity entanglement between remote nodes is challenging due to the propagation loss of photonic carriers and the imperfections and noise within local quantum nodes.

Quantum repeaters (QR) is developed to generate and maintain high-fidelity entanglement among quantum nodes over long distances overcoming the above challenges~\cite{Briegel1998,Sangouard2011,Azuma2023}. Despite advances in QPUs, the development of fully functional QRs is hindered by practical difficulties. For example, QRs based on superconducting QPUs are hampered by decoherence of microwave photons at room temperature, necessitating advanced quantum transducers~\cite{Niu2023,Storz2023,Han2021}. In the case of trapped ions and atoms, although the inherent optical photon-qubit interface enables entanglement between atoms via the interference and detection of emitted photons~\cite{Yu2020,Langenfeld2021,VanLeent2022,Krutyanskiy2023a,Krutyanskiy2023b}, the entanglement fidelity is limited by low collection efficiencies and  imperfections, such as qubit decoherence. Although entanglement purification~\cite{Bennett1996} allows the extraction of high-quality entanglement from multiple low-fidelity pairs, complex quantum operations and measurements among qubits are required. Additionally, balancing the trade-off between entanglement generation rate and achievable fidelity while ensuring the scalability and practicality of the repeater architecture is a key hurdle that needs to be addressed~\cite{Jiang2009,Sangouard2011,Munro2015,Muralidharan2016,Vinay2017,Pant2019}.

\begin{figure*}
\includegraphics[width=2\columnwidth]{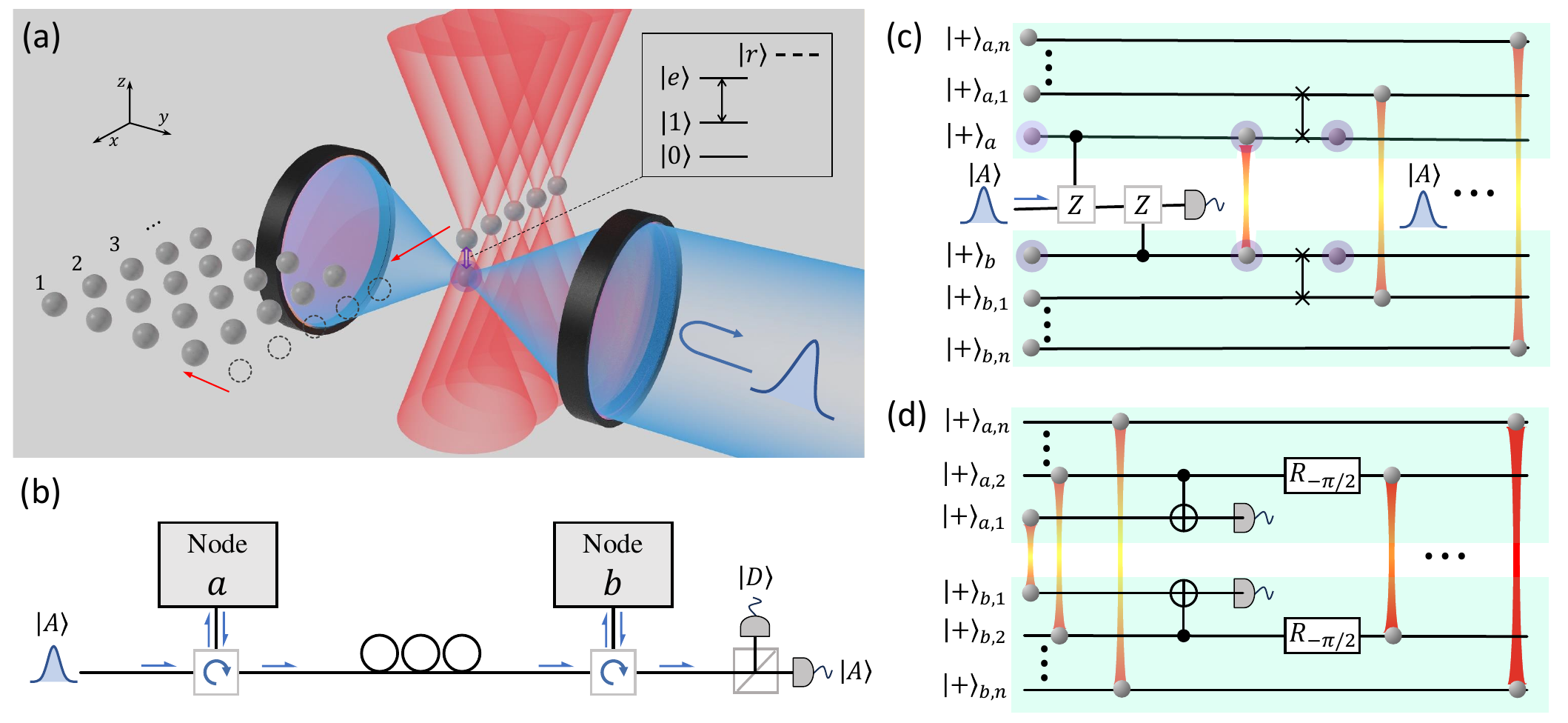}
\caption{(a) Schematic diagram of the proposed quantum repeater architecture. The initial entanglement can be generated between the communication atom trapped in the cavity mediated by ancilla photons. The entanglement is then transferred to the shuttle qubits through a Rydberg interaction-based swap gate between the communication qubit and the shuttle qubit moved with optical tweezers. By repeating this process, multiple pairs of remotely entangled shuttle qubits can be obtained. Inset: energy levels of the atom, with $\ket{0}$ and $\ket{1}$ representing the ground states used for encoding quantum information, the transition $\left|1\right\rangle \leftrightarrow\left|e\right\rangle $  to the excited state coupled to the cavity, and $\ket{r}$ the Rydberg state used for implementing gates. (b) Entanglement generation between two nodes ($a$ and $b$). An ancilla photon is sequentially couple with the atom-cavity system in each node through an optical fiber and circulators, and the remote entanglement of the communication atoms is established if a single photon is detected in the polarization sate $\left|A\right\rangle$ or $\left|D\right\rangle$. (c) The quantum circuit diagram for generating entanglement pairs between adjacent nodes. (d) The quantum circuit diagram for entanglement purification in the quantum computation zone.
}
\label{Fig1}
\end{figure*}

In this Letter, a practical QR architecture that integrates Rydberg atom arrays with cavity quantum electrodynamics (QED) is proposed. Our scheme leverages the unique strengths of Rydberg atom qubits~\cite{Saffman2010,Saffman2019}, which exhibit fast and high-fidelity quantum operations, and the exceptional atom-photon interface afforded by cavity QED systems~\cite{Wilk2007,Ritter2012,Reiserer2014,Daiss2021}. Compared with the schemes based on Rydberg-blockade-coupled atomic ensembles~\cite{Zhao2010,Han2010}, the single-atom array constituting QPUs offers unprecedented scalability and higher operation fidelities. Detailed numerical simulations, taking into account realistic experimental parameters and imperfections, demonstrate the generation of entangled pairs with fidelities exceeding $99\%$ at rates of 1.1\,kHz in two adjacent nodes. Our architecture opens a route towards large-scale quantum networks in the near term.

\textit{Architecture.-} The proposed QR architecture utilizes photons as ancillary qubits to establish remote entanglement between two nodes, leveraging the ability of photons to travel long distances through optical fiber networks. As illustrated in Fig.~\ref{Fig1}(a), each Rydberg QR node consists of several key components:

(i) \textit{Qubit-photon interface:} The architecture incorporates an atom-photon interface where an input photon with a specific polarization can become entangled with the communication qubit. Based on the Duan-Kimble scheme~\cite{Duan2004}, each node features a single-sided Fabry-Perot cavity containing a fixed atom trapped in an optical standing-wave trap potential formed by the cavity. The atom can couple with the cavity modes, enabling the quantum interface between stationary communication qubits and flying photons.

(ii) \textit{Communication qubits:} The fixed atom trapped within the cavity serves as the communication qubits, with its $\ket{1}\leftrightarrow
\ket{e}$ transition strongly coupled to a cavity mode. The inset of Fig.~\ref{Fig1}(a) depicts the energy levels of the atoms, with the ground states $\left|0\right\rangle$ and $\left|1\right\rangle$ used for encoding quantum information. The transition $\left|1\right\rangle \leftrightarrow\left|e\right\rangle$ is strongly coupled to the cavity.

(iii) \textit{Shuttle qubits:} An one-dimensional array of shuttle qubits is designed to move across the cavity without interacting with the cavity mode or the cavity dipole trap field. When a shuttle qubit travels through the cavity field, a swap gate can be realized between the communication qubit and the shuttle qubit via the Rydberg state $\ket{r}$. The shuttle qubit, capable of transporting quantum information, provide the resource of remote maximally entangled qubits (e-bits) for further quantum information processing tasks.

(iv) \textit{Local quantum computation (QC) zone:} This part serves as the core for the local QPU. Each node features a two-dimensional array of atoms acting as data qubits for implementing complex quantum tasks within the node, including addressed quantum gates, mid-circuit measurement, and feedforward operations. The atoms in the QC zone can be held either in a static tweezer array or a dynamically reconfigurable tweezer array for better qubit connectivity.

Each repeater node functions as a fully-fledged QPU, equipped with quantum communication channels that enable the interaction between the QPU and flying photons. As shown by Fig.~\ref{Fig1}(b), a probe photon is sent through the two nodes sequentially to generate e-bits between two nodes~\cite{Daiss2021}. The prepared e-bits are then transferred to the data qubits in the QC zone and stored as resources. Figure~\ref{Fig1}(c) provides a detailed quantum circuit description of the entanglement generation process between two remote Rydberg QPU nodes. The entire operation can be divided into two critical stages: e-bit generation and e-bit purification.

\textit{E-bit generation.-} The first step is to establish entanglement between the communication atoms in the cavities of the two adjacent nodes. For the atom-photon interaction, the cavity mode is decoupled from the atom if the atom is initialized to $\ket{0}$, and the system reaches the strong coupling regime when the atom is in $\ket{1}$. As a results, the reflected photon from a repeater node gains a conditional $\pi$ phase determined by the quantum state of the communication qubits~\cite{Duan2004}. By sending a probe photon sequentially through the cavities in two nodes, the communication atoms are projected into different parity states. Depending on the output photon state, the system is projected into the Bell state $\ket{\Phi^+}=\left(\left|0\right\rangle _{a}\left|0\right\rangle _{b}+\left|1\right\rangle _{a}\left|1\right\rangle _{b}\right)/\sqrt{2}$ or $\ket{\Psi^+}=\left(\left|1\right\rangle _{a}\left|0\right\rangle _{b}+\left|0\right\rangle _{a}\left|1\right\rangle _{b}\right)/\sqrt{2}$. Then, by a conditional X gate applied to node $a$, the desired e-bit $\ket{\Psi^+}$ stored in the communication qubits can be prepared.

Consider practical systems based on $^{87}\mathrm{Rb}$ atoms, various imperfections must be taken into account, such as intrinsic photon loss in the cavity, spontaneous decay of $\ket{e}$, photon propagation loss, and single photon detector errors. Based on previous experimental results~\cite{Daiss2021,Liu2023}, we set the intrinsic ($\kappa_{0}$) and total ($\kappa$) amplitude decay rates of the cavity to $2\pi\times0.2\,\mathrm{MHz}$ and $2\pi\times 4\,\mathrm{MHz}$, respectively, with the atomic amplitude decay rate $\gamma=2\pi\times3\,\mathrm{MHz}$ and the cavity-atom coupling strength $g=2\pi\times7.6\,\mathrm{MHz}$. The external coupling rate of the cavity, $\kappa_{\mathrm{ex}}=\kappa-\kappa_0$, is specifically designed to achieve a unit post-selected photon-atom controlled-Z (CZ) gate fidelity, with a success probability of $81\%$~\cite{SM}. Considering the finite bandwidth of the cavity, the probe photon pulse length is set to $20/\kappa$ to ensure the steady-state assumption. For two nodes connected by a $l=100\,\mathrm{m}$ fiber (propagation loss $3\,\mathrm{dB/km}$ for $780\,\mathrm{nm}$), and accounting for insertion losses of fiber circulators ($1\,\mathrm{dB}$) and imperfect detection efficiencies ($75\%$), the overall success probability of generating entanglement between communication atoms is $P_{\mathrm{succ}}=36\%$. Although post-selection can mitigate CZ gate imperfections, technical imperfections still contribute to e-bit infidelities~\cite{Daiss2021}, including atom state initialization error ($0.01$), cavity length fluctuation ($0.01$), fiber depolarization ($0.01$), and mode matching at the beam splitter ($0.01$). Note that  single qubit gate error is negligible~\cite{Sheng2018}. Considering these factors, an e-bit fidelity of $F\approx0.96$ is experimentally feasible. Ignoring the time required for initialization and single-photon detection ($\ll1\,\mathrm{\mu s}$), the expected time to prepare one e-bit is $T_{\mathrm{esta}}=\left({20}/{\kappa}+{l}/{v}+{l}/{c}\right)/P_{\mathrm{succ}}\approx4.53\,\mathrm{\mu s}$, where $v$ is the speed of light in the fiber. This results in an e-bit generation rate of $1/T_{\mathrm{esta}}\approx221\,\mathrm{kHz}$~\cite{SM}.

\textit{E-bit purification.-} Once an e-bit is generated between the communication qubits in two adjacent nodes, it is transferred to a nearby shuttle qubit through a local swap gate, which is implemented by a sequence of three Rydberg-state-mediated controlled-NOT (CNOT) gates. According to the error model of two-qubit gates from Ref.~\cite{Evered2023}, the estimated fidelity of CNOT gate is $F_{op}=0.995$~\cite{Evered2023,Radnaev2024}, give rise to a fidelity of SWAP operation exceeding $0.98$ with a duration of $T_{\mathrm{swap}}=2\,\mathrm{\mu s}$ (see Ref.~\cite{SM} for details). A chain of shuttle qubits is queued up for e-bit preparation. After the swap gate, the shuttle qubit storing the e-bit is transported out, and a fresh shuttle qubit is brought in to approach the communication qubit for preparing the next e-bit. The queue of shuttle qubits storing the e-bit are moved to the QC zone for further applications. At a speed of $0.55\,\mathrm{\mu m/\mu s}$~\cite{Bluvstein2024,bluvstein2022quantum}, the fidelity of the e-bit is well preserved, exceeding $0.96$ for an average duration of $T_{\mathrm{move}}=20\,\mathrm{\mu s}$, with a transportation successful probability is $P_{\mathrm{succ}}\geq90\%$~\cite{SM}.

Although initial e-bit fidelities are limited by imperfections in the preparation and transportation processes, the quality of the e-bits can be enhanced through entanglement purification operations while consuming multiple copy of low-quality e-bits. Figure~\ref{Fig1}(d) illustrates the quantum circuit diagram of the entanglement purification, which can be decomposed into two CNOT gates and detection operations. The operation fidelity for purification considers the combination of errors in CNOT gate ($0.005$) and detection ($0.01$) on both nodes, and also the successful probability $P_{\mathrm{puri}}$. The duration of the purification is primarily limited by the readout time of the atom states $T_{\mathrm{proj}}$, the average time cost for a purification step is given by $T_{\mathrm{puri}}=T_{\mathrm{proj}}/P_{\mathrm{puri}}$. Employing a single-photon detector, high-fidelity detection of single atom can be achieved with a duration of $T_{\mathrm{puri}}<400\,\mathrm{\mu s}$~\cite{Shea2020,Chow2023}.

\renewcommand{\tablename}{\textbf{Table}}
\begin{table}[b]
    \setlength{\tabcolsep}{1.2mm}{
    \caption{\textbf{The duration, fidelity, and success probability ($P_{\mathrm{succ}}$) for the operations.}}
    \centering
    \begin{tabular}{lcccc}

    \hline
   \textbf{} & \textbf{Entanglement} & \textbf{Swap} &  \textbf{Move} & \textbf{Purification}\tabularnewline

   \hline
    \hline
    Duration ($\mathrm{\mu s}$) & 3.6& 2& 20 & $<400$\tabularnewline

    Fidelity & 0.96 & 0.98& 0.96 & $\sim 0.99$\tabularnewline

    $P_{\mathrm{succ}}$ & 36\% & / & 90\% &  $>50\%$ \tabularnewline
    \hline
\label{tab}
\end{tabular}
}
\end{table}

\begin{figure}
\includegraphics[width=\columnwidth]{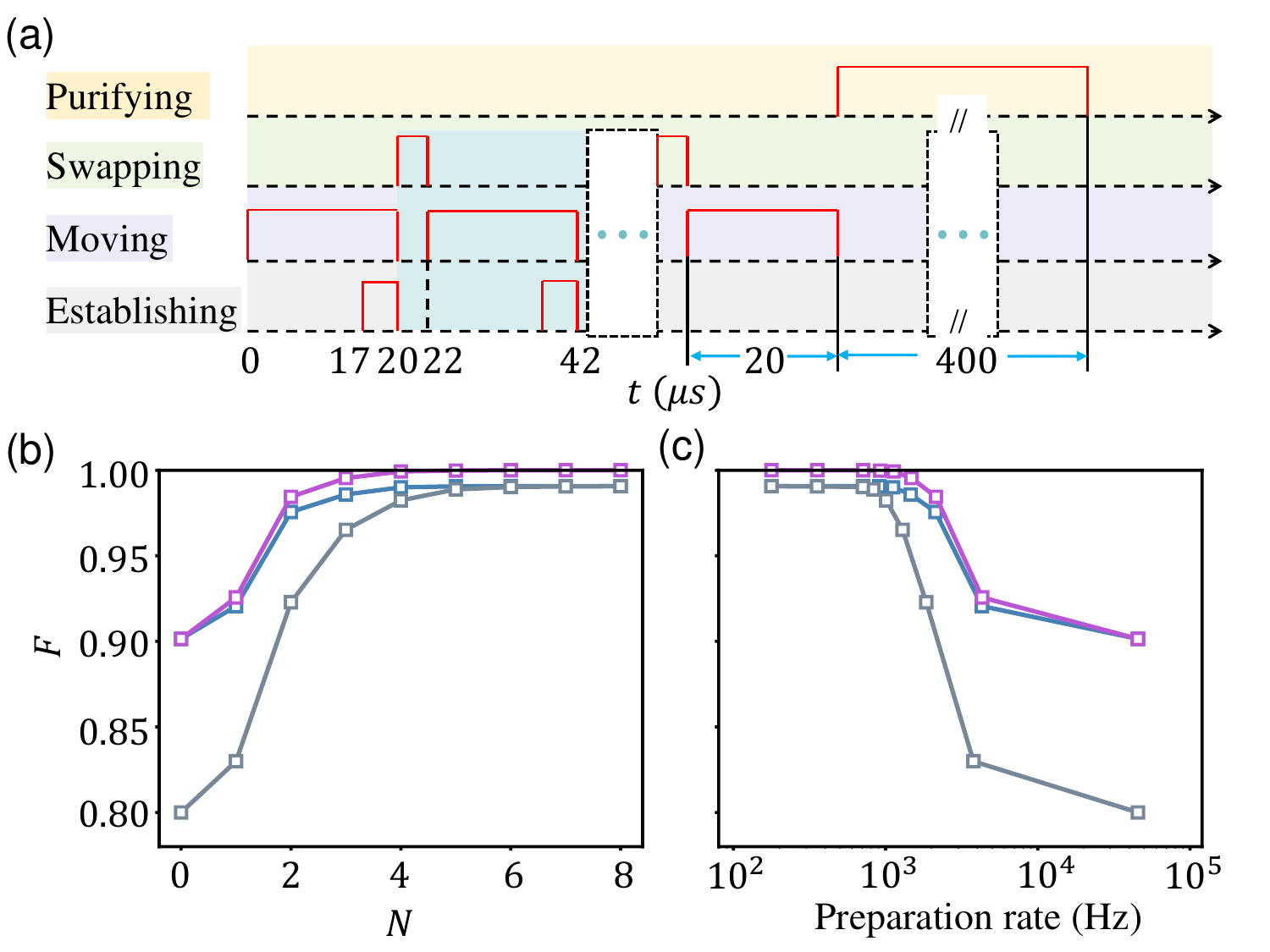}
\caption{(a) The optimized pulse sequences  for the full entanglement generation and purification process, including establishing initial entanglement, moving atoms, swap gate, and entanglement purification. Entanglement generation and purification can be carried out in parallel to maximize efficiency. (b) and (c) The fidelity and effective generation rate of the purified e-bits as a function of the number of purification rounds, $N$. The purple line represents the ideal purification with perfect operations and an initial fidelity of $F_{0}\approx0.91$. The blue and gray lines show the results for imperfect operations (CNOT gate fidelity of 0.995 and detection accuracy of 0.99) with initial fidelities of $F_{0}\approx0.91$ and $0.8$, respectively. }
\label{Fig2}
\end{figure}

Table~\ref{tab} summarizes the duration, fidelity and success probability of essential individual operations. The eventual generation rate and fidelity of e-bit are determined by the execution sequences of these operations (as illustrated by Fig.~\ref{Fig2}(a)) and the rounds of the purification ($N$). Since the hyperfine energy levels of atoms can work with a coherence time for seconds~\cite{Tian2023}, we neglects qubit decoherence when idling the the optical tweezers. According to Table~\ref{tab}, the initial e-bit fidelity in the QC zone is limited to around $0.90$. Figure~\ref{Fig2}(b) shows the numerical results of $N$-round purification, which generates high-quality e-bits by consuming $2^N$ lower-quality e-bits. As shown by the blue line, the e-bit fidelity quickly converges to $F\sim0.99$ when $N\geq4$ with experimentally feasible parameters. Even for a lower initial fidelity of $0.8$ (gray line), the fidelity can be converged to the same value when $N\geq6$. To reveal the effects of imperfect gate operations, we plot results with perfect purification operations (red line) as comparison.

The generation and processing of e-bits in our architecture can be optimized by treating the operations as an assembly line, where the tasks are synchronized for maximum efficiency. In particular, the operations on the cavity, shuttle atoms, and QC zone can be implemented in parallel. The total time consumed by preparing $2^N$ e-bits and implementing $N$ rounds of purification is estimated as$T_{EG,N}=\mathrm{max}\left\{ 2^N\left[\mathrm{max}\left(T_{\mathrm{esta}}+T_{\mathrm{swap}},T_{\mathrm{swap}}+T_{\mathrm{move}}\right)\right], N\left(T_{\mathrm{puri}}+l/c\right)\right\}$, where $l/c$ is the time required for classical communication. Taking into account the overall success probability, the relationship between effective e-bit generation rate and fidelity after purification is summarized in Fig.~\ref{Fig2}(c). Without purification ($N=0$), the e-bit with $F=0.9$ in the QC zone can be produced with an effective rate of $R=45\,\mathrm{kHz}$. For $N=4$, e-bit with $F=0.99$ can be generated at a rate of  $R=1.1\,\mathrm{kHz}$.

\begin{figure}
\includegraphics[width=\columnwidth]{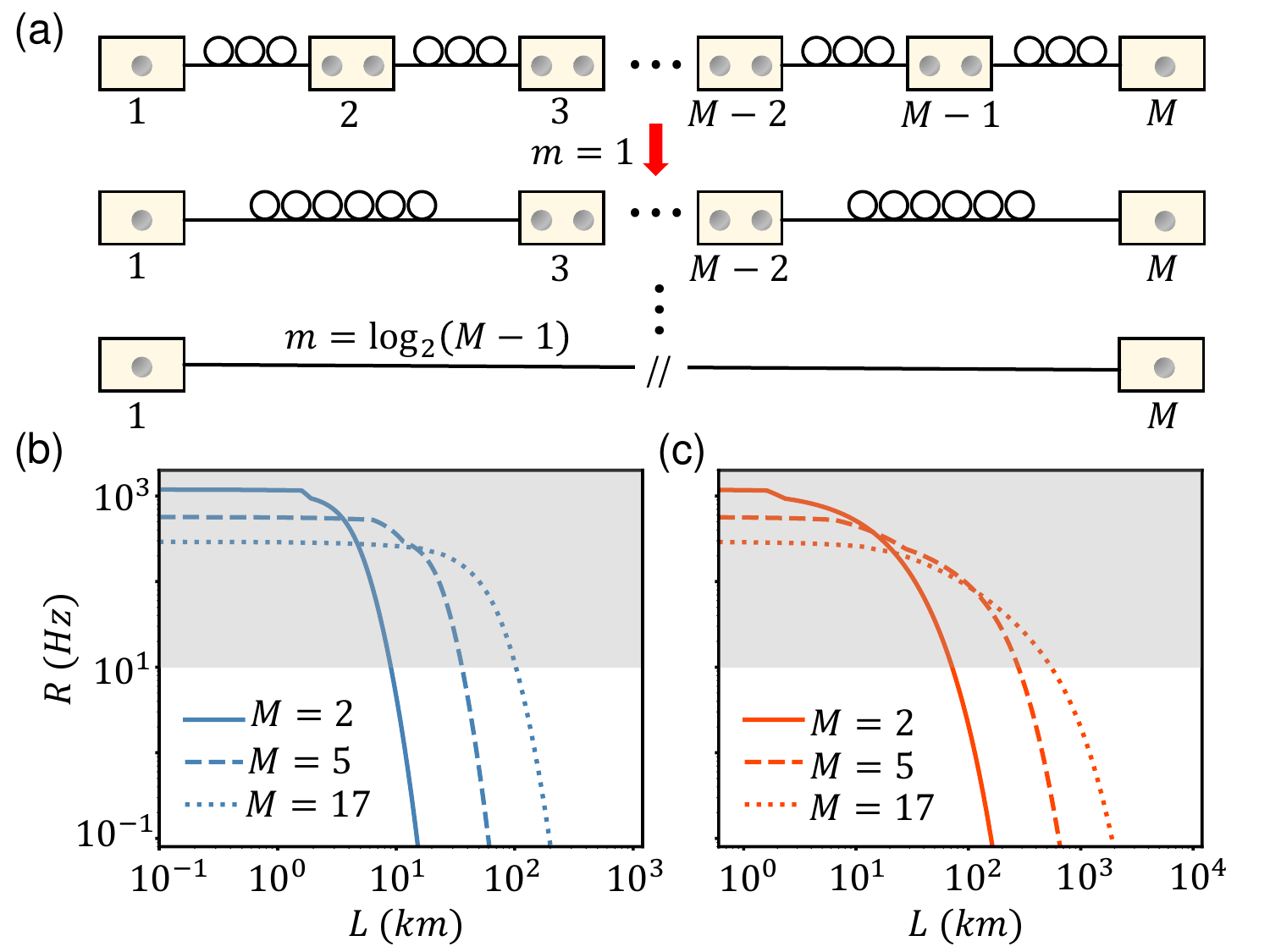}
\caption{ (a) Schematic of the entanglement distribution process in a quantum repeater network. The remote entanglement between the end nodes of a chain of $M$ repeater stations is generated through parallel entanglement swapping operations, with $m=\log_{2}(M-1)$ iteration steps. (b) The achievable entanglement generation rate as a function of the total distance $L$ between the end nodes, for different numbers of repeater stations ($M=2$, $5$, and $17$) with a fidelity of e-bit $F_{M}\geq0.99$. (C) The entanglement generation rate as a function of the total distance $L$ for $M=2$, $5$, and $17$ with $F_{M}\geq0.99$, with the introduction of frequency conversion from 780\,nm to the telecommunication wavelength of 1550\,nm. }
\label{Fig3}
\end{figure}

\textit{Quantum repeater network.-} Taking the advantage of overcoming the propagation loss and operation errors when build entanglement between two nodes, the distance of entangled atoms can be effective extended with each nodes serving as QR station, as shown in Fig.~\ref{Fig3}(a). For a chain of $M$ stations, remote entanglement with a distance of $L=(M-1)l$ can be generated by implementing $m=\log_{2}\left(M-1\right)$ parallel entanglement swapping operations between the stations. For example, by performing a Bell measurement on a pair of e-bits in station 2, the entanglement between stations 1 and 3 can be generated, followed by a feedforward operation. The fidelity of the generated e-bit for the two ends of the chain is simulated by the quantum channel $\epsilon_{CZ}(\rho)$ with fideity $F_{op}$ and POVM $\mathcal{M}(\rho)$ with error rate $\eta$. Consistent with the parameters in Table~\ref{tab}, we set $F_{\mathrm{op}}\approx0.995$ and $\eta=0.99$.  All the Bell measurement for entanglement swapping can be done together,  and each node sends the result of Bell measurement to the closer node among the beginning and final nodes,  followed by feedforward operations for nodes at both ends, so the consumed time for the QR operations is $T_{\mathrm{repe}}\approx L/(2c)+  T_{\mathrm{proj}}$, where $L/(2c)$ is the whole time for the classical communication. We perform purification before and after the entanglement swapping with $N_1$ steps and $N_2$ steps, respectively. Then we optimize $N_1$ and $N_2$ to get the e-bit fidelity $F_M \geq F$ and a optimized total time. Therefore, the total time consumed from establishing entanglement to the QR is $T_{QR}=\mathrm{min}_{N_1,\,N_2}\{T_{EG,N_1}+T_{repe}+T_{EG,N_2}\}|_{F_M\geq F}$.

Figure~\ref{Fig3}(b) plots e-bit preparation rate as a function of the distance $L$ for different number of QR stations ($M=2$, $5$ and $17$), with initial e-bit fidelity $F_{M}\geq0.99$. The numerical results indicate that the proposed QR based on $780\,\mathrm{nm}$ photons supports a $100\,\mathrm{Hz}$-level high-fidelity e-bit generation rate for the distance to metropolitan area ($L\sim25\,\mathrm{km}$) for only $M=5$ stations. When extending to $L > 100 \,\mathrm{km}$, the rate is reduced to near $10.0\,\mathrm{Hz}$. To extend the entanglement to even longer distance, such as for realizing intercity quantum network, the transmission loss of photons in the fiber can be suppressed by converting the photons to telecommunication wavelength $1550\,\mathrm{nm}$ ($0.19\,\mathrm{dB/km}$)~\cite{VanLeent2020,Liu2024}. Assuming a conversion efficiency of $\eta_{\mathrm{FC}}=0.6$, an additional $\eta_{\mathrm{FC}}^2$ insertion loss is introduced when generating e-bit between adjacent nodes. The corresponding e-bits generation rate against $L$ is plotted in Fig.~\ref{Fig3}(c). The introduction of frequency conversion further extends remote entanglement to $250\,\mathrm{km}$ with a high-fidelity e-bit generation rate at the 10 Hz level, requiring only $M\geq5$ QR stations. When employ $M=17$ stations, high fidelity (0.99) e-bits can be generated at a rate of $10\,\mathrm{Hz}$ even for $L>500\,\mathrm{km}$.

\textit{Conclusion.-} The proposed QR architecture based on Rydberg atom arrays and cavity-QED systems demonstrates promising potential for achieving long-distance entanglement distribution. QR schemes can be classified into three generations based on their treatment of photon loss errors and operation errors~\cite{Muralidharan2016,Azuma2023}. Our scheme belongs to the first generation, where both types of errors are suppressed via post-selection. There are several aspects to improve the e-bit generation rate in QR networks: extend the system to multiple cavities to generate e-bits in parallel, and optimize the sequences of purification and entanglement swapping in the network~\cite{Pant2019}. By using Alkali-earth atoms and  erasure conversion can improve the e-bit fidelity after postselection \cite{Ma2023-oh, Wu2022-xq}. Our architecture also offers the possibility of implementing quantum error correction encoding based on data qubits in the QC zone. This feature makes our architecture compatible with the second-generation QR schemes~\cite{Jiang2009,Azuma2023}, which is worth further investigation for achieving even higher e-bit generation rates.

\begin{acknowledgments}
This work was funded by the National Key R\&D Program (Grant No.~2021YFA1402004), the National Natural Science Foundation of China (Grants No. U21A20433, U21A6006, 92265210, and 92265108), and Innovation Program for Quantum Science and Technology (Grant No.~2021ZD0300203). This work was also supported by the Fundamental Research Funds for the Central Universities and USTC Research Funds of the Double First-Class Initiative. The numerical calculations in this paper have been done on the supercomputing system in the Supercomputing Center of University of Science and Technology of China. This work was partially carried out at the USTC Center for Micro and Nanoscale Research and Fabrication.
\end{acknowledgments}

\end{document}